# Full-Spectrum Wireless Communications for 6G and Beyond: From Microwave, Millimeter-Wave, Terahertz to Lightwave

Wei Jiang
Intelligent Networking Group
German Research Center for AI
(DFKI)
Kaiserslautern, Germany

Hans D. Schotten
*Institute for Wireless
Communication and Navigation*
RPTU Kaiserslautern-Landau
Kaiserslautern, Germany

*Abstract*—As of today, 5G is rolling out across the world, but academia and industry have shifted their attention to the sixth generation (6G) cellular technology for a full-digitalized, intelligent society in 2030 and beyond. 6G demands far more bandwidth to support extreme performance, exacerbating the problem of spectrum shortage in mobile communications. In this context, this paper proposes a novel concept coined Full-Spectrum Wireless Communications (FSWC). It makes use of all communication-feasible spectral resources over the whole electromagnetic (EW) spectrum, from microwave, millimeter wave, terahertz (THz), infrared light, visible light, to ultraviolet light. FSWC not only provides sufficient bandwidth but also enables new paradigms taking advantage of peculiarities on different EW bands. This paper will define FSWC, justify its necessity for 6G, and then discuss the opportunities and challenges of exploiting THz and optical bands.

## I. Introduction

When South Korea and USA were competing with one another for the world's first Fifth-Generation (5G) commercial deployment in April 2019, we stepped into the era of 5G. As of today, 5G availability is realized in most countries and territories worldwide, and the penetration rate of 5G usage expands quickly. For example, China has deployed over 2 million 5G base stations to serve approximately one billion 5G subscribers at the end of 2022.

Although 5G still needs a long journey to reach its full potential in reshaping verticals and society as expected, academia and industry have already shifted their focus towards the next-generation technology known as the Sixth-Generation (6G) mobile communications [1]. Leading countries in the wireless-communications industry have initiated their efforts, aiming to get a leading position in the race of developing 6G. As early as 2018, Finland pioneered this competition by starting the world's first 6G program *6G-Enabled Wireless Smart Society and Ecosystem (6Genesis)* at the University of Oulu [2]. In November 2019, the Ministry of Science and Technology of China has officially announced the national 6G program, and meanwhile established a promotion working group and an overall expert group. Under its *Horizon 2020* program (FP8), European Commission kicked off its 6G research in early 2020 through ICT-20-2019 call *5G Long Term Evolution*, followed by another batch projects - ICT-52-2020 *Smart Connectivity beyond 5G* - at the beginning of 2021. In October 2020, the Alliance for Telecommunications Industry Solutions (ATIS) launched *Next G Alliance* [3] with founding industrial members such as AT&T, T-Mobile, Verizon, Qualcomm, Ericsson, Nokia, Apple, Google, Facebook, Microsoft, aiming to advance U.S. leadership in 6G. In August 2021, the German Federal Ministry of Education and Research provided a total fund of 250 million euros, as the first four-year phase, to establish four 6G research hubs, i.e., 6G-life, 6GEM, 6G RIC, and Open6GHub [4]. Japan dedicates $2 billion to foster the private sector for developing 6G. South Korea invested $169 million over the first five years to develop 6G technologies, intending to conduct the first 6G trial worldwide in 2026.

In 2018, ITU-T set up a focus group *Technologies for Network 2030* to identify the technological gaps for the capacities of networks in 2030 and beyond. It published a technical report, which envisages that 6G will support disruptive applications, such as holographic telepresence, virtual reality, autonomous driving, Tactile Internet, ubiquitous intelligence, and digital twin [5]. These novel uses impose extreme capacity and performance requirements on 6G [6], e.g., a peak data rate of 1 terabits-per-second (Tbps), a massive connection density of 10,000,000 devices per square kilometer, mobility support for up to 1,000 kilometers per hour, centimeter-level positioning accuracy, and an area traffic capacity of 1Gbps per square meter. In addition to innovative transmission and networking technologies, 6G needs far more bandwidth that never encountered in the previous generations to fully implement these key performance indicators (KPIs). This will further exacerbate the problem of spectrum shortage in mobile communications. The next section elaborates on the spectrum availability in current systems and justify the necessity of exploiting higher frequency bands.

## II. Evolution and Spectrum Shortage in Mobile Communications

For the past decades, the evolution of mobile communications followed these important rules:

- **Signal bandwidth becomes increasingly wide;**
- **Operating frequency band is increasingly high;**
- **Spectral demand is increasingly large**.

We can see that a new system demands more spectral resources and utilizes a larger channel bandwidth to provide more system capacity and realize a higher data rate than its predecessor [7]. Initially, the bandwidth of each signal channel in the First-Generation (1G) system is only 20~30KHz that is already sufficient to carry the analog voice signal of a mobile user. For example, the most dominating 1G standard - Advanced Mobile Phone System (AMPS) - utilizes a pair of 25MHz (for downlink and uplink, respectively) to carry maximally 832 analog voice users at each cell cluster. At that time, the low-frequency band under 1GHz with favorable propagation characteristics is a preferred choice for system designers.

Global System for Mobile communications (GSM) first launched in 1991 supports 1,000 subscribers within 2x25MHz bandwidth, where eight digital-voice users are multiplexed over each 200KHz channel using time-division multiple access (TDMA). For the Third Generation (3G), Wideband Code-Division Multiple Access (WCDMA) employs a much wider bandwidth of 5MHz to carry tens of users per channel simultaneously. Meanwhile, the frequency band crossed over 2GHz for the first time due to the demand of more spectral resources. Further, Long-Term Evolution-Advanced (LTE-A), as the unified Fourth-Generation (4G) standard, supports a maximal bandwidth of 100MHz using carrier aggregation so as to realize the peak rate of 1Gbps in the downlink. Accordingly, its frequency band spans over a wider range, from 450MHz to 6GHz, since hundreds of MHz spectral resources are needed to satisfy the explosive traffic growth of mobile Internet. Until 4G, the cellular systems operate in low-frequency bands below 6GHz, which are referred to as sub-6GHz band when high-frequency bands are considered in 5G New Radio (NR). A summary of previous mobile generations with the emphasis on bandwidth and operating frequency bands is given in Table I.

TABLE I. SUMMARY OF MOBILE GENERATIONS

|  | Mobile Generations | | | | |
|---|---|---|---|---|---|
|  | *1G* | *2G* | *3G* | *4G* | *5G* |
| Major Standard | **AMPS** | **GSM** | **WCDMA** | **LTE-A** | **NR** |
| Launch Year | 1979 | 1991 | 2000 | 2009 | 2019 |
| Peak Rate | N/A | 384Kbps | 2Mbps | 1Gbps | 20Gbps |
| Bandwidth | 30KHz | 200KHz | 5MHz | 100MHz | 1GHz |
| Frequency Band | 800MHz | 800MHz/ 1900MHz | Below 2.1GHz | Sub-6GHz | Sub-6GHz /mmWave |
| PHY | FDMA | TDMA | CDMA | OFDMA | OFDMA/ NOMA |

5G NR supports a peak rate of 20Gbps, where the maximal bandwidth reaches 1GHz, imposing the need of exploiting higher frequencies. That is because the sub-6G band is already overcrowded. The total amount of spectral resources assigned to IMT services in many regions is usually less than 1GHz, while finding a large-volume contiguous bandwidth below 6GHz is infeasible. In contrast, higher frequencies still have a large number of available spectra at which were already used for a wide range of non-cellular uses such as satellite communications, remote sensing, radio astronomy, radar, to name a few. With the progress in antenna and RF technology, mmWave previously considered unsuitable for cellular communications due to their unfavorable propagation features becomes technologically usable. As a result, 5G became the first cellular system to operate in the mmWave band [8].

In 2015, an item at the ITU World Radio Conference (WRC-15) was formed to study the usability of high-frequency bands above 24GHz for International Mobile Telecommunications 2020 (IMT-2020) standard. According to its results, World Radio Conference held in 2019 (WRC-19) made a note that ultra-low latency, high-data-rate applications need larger contiguous spectrum blocks. Consequently, a total of 13.5GHz spectrum consisting of a set of mmWave bands were assigned for 5G, i.e., 24.25~27.5GHz, 37~43.5GHz, 45.5~47GHz, 47.2~48.2GHz, and 66~71GHz. That motivates 3GPP to specify 24.25GHz to 52.6GHz as the Second Frequency Range (FR2) for operating 5G NR at mmWave.

### III. FULL-SPECTRUM WIRELESS COMMUNICATIONS

Nevertheless, the assigned spectral resource for IMT services is still very limited compared to the demand of 6G. Assume a total bandwidth of 10GHz, Tbps data transmission is possible merely with extremely high spectral efficiency approaching 100bps/Hz, which is not feasible under currently known techniques and hardware. Guided by the Shannon theorem, i.e., C=Blog(1+SNR), 6G has to acquire more bandwidth to achieve Tbps and other stringent performance. The only solution is to utilize the massively abundant spectrum over 100GHz, including the THz band covering 0.1~10THz and the optical band. In 2019, ITU-R WRC-19 has recognized the open of 275GHz~450GHz for IMT services, paving the way of deploying 6G THz commutations from the perspective of regulation [9]. On the other hand, much higher frequencies, namely the optical band including infrared, visible light, and ultraviolet are also friendly for wireless communications without licensing permission from regulators. Recently, optical wireless communications (OWC) attracted a lot of attention, which will be also a potential enabler for 6G [10].

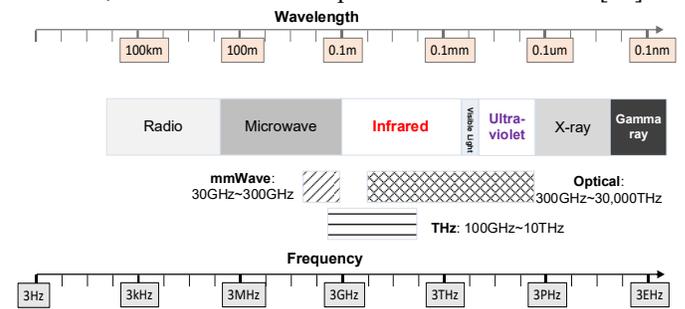

Fig. 1. The electromagnetic spectrum and the positions of mmWave, THz, and optical bands.

Exploiting all communication-friendly bands over the whole electromagnetic (EW) spectrum, this paper proposes a novel concept coined Full-Spectrum Wireless Communications

(FSWC). The whole EW spectrum is shown as Fig. 1, and the major characteristics of different frequency bands are listed in Table II. It is noted that the definition of the EW spectrum in the general case is different with the naming of frequency bands from the perspective of mobile communications. The EW spectrum is divided into radio, microwave, infrared, visible light, ultraviolet, X-ray, and Gamma ray, from the lower to higher frequencies. The so-called mmWave and THz bands are a portion of these EW bands.

TABLE II.    LIST OF EW BANDS AND THEIR APPLICATIONS

| Band | Frequency | Wavelength | Applications |
|---|---|---|---|
| Radio | 1m~100km | 3KHz~300MHz | AM/FM broadcasting<br>Radio navigation<br>Radiolocation<br>Radio astronomy<br>Radar, Remote sensing |
| Microwave | 1mm~1m | 0.3~300GHz | Satellite communications<br>Mobile communications<br>Navigation<br>Heating<br>Spectroscopy<br>Radio astronomy<br>Radar, Remote sensing |
| Infrared | 0.75um~1mm | 300GHz~400THz | Short-range communications<br>Night vision, Heating<br>Thermography, Spectroscopy<br>Astronomy, Tracking<br>Hyperspectral imaging |
| Visible light | 0.38um~0.75um | 400~790THz | Illumination<br>Heating<br>Power generation<br>Biological systems<br>Spectroscopy |
| Ultraviolet | 10nm~0.38um | 790~30PHz | Semiconductor manufacturing<br>Material science<br>Photography<br>Biological use |
| X-Ray | 10pm~10nm | 30PHz~30EHz | Medicine<br>Radiography, Radiotherapy<br>Fluoroscopy<br>Computed tomography |
| Gamma ray | ~10nm | 30EHz~ | Medicine<br>Radiography<br>Nuclear industry |

Since the particles of high-end ultraviolet light, X-rays, and Gamma rays have sufficiently high energy to dislodge electrons and create free radicals that lead to cancer, the extreme high frequency bands bring ionizing radiation. The adverse health effects of ionizing radiation may be negligible if used with care, but it is still dangerous for cellular communications. Unlike ionizing radiation, mmWave, THz, and optical waves are non-ionizing because the photon energy is not nearly sufficient to release an electron from an atom or a molecule, where typically 12 eV is required for ionization. Since ionizing radiation is not determined to be a concern at mmWave, THz, and optical bands, heating might be the only primary cancer risk. The International Commission on Non-Ionizing Radiation Protection (ICNIRP) standards have been set up to protect against thermal hazards, particularly for the eyes and skin, which are most sensitive to heat from radiation due to lack of blood flow.

Reviewing all EW bands, we can then define Full-Spectrum Wireless Communications as a wireless communications system that flexibly operates in all communication-feasible frequency bands, including radio wave, microwave, mmWave, THz, infrared, visible light, and ultraviolet, and takes advantage of peculiarities on different EW bands for specific objectives. It not only provides sufficient bandwidth for the realization of 6G performance but also enables novel networking paradigms such as integrated communication and sensing, intra-building optical backhaul, and inter-satellite link.

Since the research community already has profound knowledge and expertise on microwave and mmWave, the emphasis of the remainder of this paper is on the opportunities and challenges of THz and optical bands for 6G. Let's take a look at the synergy of different bands in FSWC, and clarify the opportunities enabled by THz and optical bands:

- Radio band is the golden spectrum for long-range coverage due to large propagation distance and high penetration capability. For example, IEEE 802.22 [11], a standard for Wireless Regional Area Network (WRAN) using TV white spaces of 54~862 MHz, achieves a maximal cell radius of 100km. This band is helpful for ubiquitous services in 6G, e.g., wide-coverage Internet-of-Things, Low Power Wide Area Network (LPWAN), public safety and emergency communications, and universal service for rural and remote areas.
- Microwave band, which is the anchor for the previous generations of mobile communications, can be reclaimed for deploying 6G macro and micro base stations to provide enhanced mobile broadband services.
- Millimeter wave is applied to alleviate the spectrum shortage problem of 6G for small cells in urban areas, indoor hot spots, high-capacity backhauling, and integrated backhaul and access, etc.
- Super-wide THz bandwidth facilitates a variety of ultra-high-rate 6G uses, e.g., holographic telepresence, fully immersive experience, ultra-low-latency video delivery without compression, over-the-air AI computing, information shower, and Tactile Internet. It empowers another degree of freedom for designing a flexible mobile network. For example, THz wireless backhaul can speed up system deployment and reduce maintenance cost. THz wavelength is tiny, paving the way of developing revolutionary applications such as nanoscale networks, on-chip communications, Internet of Nano-Things, and intra-body network. It can also be combined with biocompatible and energy-efficient nanodevices to realize molecular communication using chemical signals. THz is also promising for efficiently implementing integrated communications and sensing since tiny wavelength helps high-accuracy sensing, imaging, and positioning of the surrounding physical environment.
- Optical band has almost unlimited, license-free bandwidth. Given off-the-shelf optical emitters and detectors (i.e., light-emitting diodes, laser diodes, and photodiodes), it provides a low-cost solution for 6G. Since lightwave is immune to RF interference and does not bring radio radiation to humans, it is favorable for some deployment scenarios such as home networking, vehicular communications, airplane passenger cabin, and hospitals with interference-sensitive medical equipment. Since optical beams are highly directive and is not penetrable to blockage, it is an excellent medium for security uses. Relying on high-power, high-concentrated laser beams,

Free-Space Optical (FSO) communications achieve high data rate over a long distance up to several thousand kilometers. It offers a cost-effective tool for crosslinks among space, air, and terrestrial platforms, and facilitates inter-satellite links for emerging LEO satellite constellations. In addition, underwater optical communications provide higher transmission rates than acoustic communications with significantly low power consumption and low complexity. In a nutshell, the optical band is critical for building ubiquitous terrestrial-air-space-ocean 6G coverage.

## IV. CHALLENGES OF THZ BAND

Despite its advantages, THz signal transmission suffers from some impairments that are not severe in low-frequency bands:

### A. High Path loss

Path loss is an inherent attenuation when an EW wave is radiated. Due to the dependence of the aperture area of a receiving antenna on wavelength, path loss proportionally grows with the increase of carrier frequency. Hence, the THz wave suffers from much higher loss since the receiving THz antenna is tiny and has a weak ability to capture the radiation power. This loss increases 20dB per decade as a function of carrier frequencies. For example, it gets an extra loss of 20dB from 30GHz to 300GHz under the same condition.

### B. Gaseous Attenuation

Although gaseous molecules absorb some energy of an EW wave, atmospheric absorption is small enough over the sub-6G band such that traditional cellular systems do not take it into account when calculating the link budget. However, this effect substantially magnifies for the THz wave, and it becomes extremely large at certain frequencies. When the THz wavelength approaches the size of gaseous molecules, the incident wave causes rotational and vibrational transitions in polar molecules. As a main gaseous component of the atmosphere, oxygen plays a major role in atmospheric absorption under clear air conditions. In addition, water vapor suspended in the air strongly affects the propagation of an electromagnetic wave. The attenuation caused by water vapor dominates the THz band, except only a few specific spectral regions where the effect of oxygen is more evident [12].

Given a usual atmospheric condition at the seal level with an air pressure of 1013.25hPa, the temperature of 15 degrees Celsius, and a water-vapor density of 7.5mg/m$^3$, the atmospheric absorption up to 1THz is illustrated in Fig. 2. As we can see, this loss accounts for a peak of approximately 20,000dB per kilometer in the worst case. In other words, only a short distance of 1m brings a loss of approximately 20dB, which is prohibitive for efficient wireless communications. As expected, the atmospheric attenuation at the low-frequency band is negligible.

### C. Weather Effects

Besides the gaseous absorption, an additional atmospheric impact in an outdoor environment is the weather. Extensive studies of satellite communication channels since the 1970s provided many insights into the propagation characteristics of THz signals under various weather conditions. Like water vapor in the atmosphere, the outcomes revealed that liquid water droplets, in the form of suspended particles in clouds, fogs, snowflakes, or rain falling hydrometeors, absorb or scatter the incident signals since their physical dimensions are in the same order as the THz wavelength. Fig. 3 illustrates the rain attenuation derived from ITU-R model [13]. Such attenuation is not as strong as the path loss and atmospheric absorption but still needs to be considered.

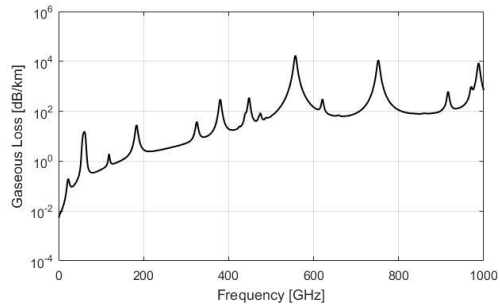

Fig. 2. Gaseous absorption in dB/km from 1GHz to 1THz under the standard atmospheric condition.

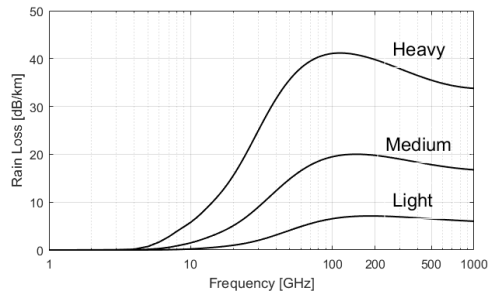

Fig. 3. Rain attenuation in dB/km from 1GHz to 1THz under different rainfall rates: heavy (150mm/h), medium (50mm/h), and light (10mm/h).

### D. Blockage

Due to the tiny wavelength in THz, the dimensions of surrounding physical objects are sufficiently large for scattering, while specular reflections on ordinary surfaces become difficult. On the other hand, THz systems rely heavily on pencil beams to extend distance. As a result, a direct path between the transmitter and receiver is desired. However, the LOS THz link is highly susceptible to being blocked by macro-objects, such as buildings, furniture, vehicles, and foliage, and humans, in comparison with the traditional sub-6G band. Such blockage losses can dramatically decay the signal power and may even lead to a thorough outage. Hence, it is necessary to find effective solutions to avoid be blocked.

## V. CHALLENGES OF OPTICAL BAND

Like the THz band, there exist some challenges for lightwave transmission.

### A. Weather Attenuation

Adverse weather conditions cause a power loss of lightwave on tens or hundreds of dBs per kilometer and distorts the transmitted signal. Unlike the THz band, raindrops and snowflakes are much larger than the wavelength of the optical band. The most harmful atmospheric conditions turn to fog and haze due to their comparable radii on the same order of magnitude as the optical wavelength [14].

*B. Atmospheric Absorption*

Like THz wave, atmospheric molecules such as oxygen and water vapor absorb the energy of light, imposing attenuation of the optical power. This absorption depends on the wavelength so that a given atmospheric situation may be transparent to some types of lights while completely blocking others. For example, infrared light is primarily absorbed by water vapor and Carbon Dioxide ($CO_2$) in the atmosphere, while absorption by oxygen ($O_2$) and ozone ($O_3$) strongly attenuate ultraviolet light [15].

*C. Turbulence*

The varying atmospheric conditions cause turbulence due to different temperatures, atmospheric pressures, and humidity levels within the propagation media. It leads to scintillation of the light and beam wander. In other words, the optical power acquired by a receiver fluctuates over time and the position of the incident light shifts in space. This phenomenon steps from air conditioning vents in the vicinity of the transceiver, irradiative heat from the roof, or currents of pollutants in the atmosphere. Temperature and humidity gradients cause changes in the atmospheric refractive index, which is the source of optical distortions. Winds and cloud coverage also influence the level of turbulence, and even the time of day can alter temperature gradients.

*D. Beam Alignment*

Outdoor optical wireless communications heavily rely on LOS transmission. Therefore, it is imperative to maintain the alignment between transmitter and receiver continuously. However, accurate alignment is particularly challenging with narrow beam divergence angles at the transmitter and narrow Fields of View (FOV) at the receiver. Outdoor OWC equipment is generally installed on the top of high buildings. Building sway due to thermal expansion of building frame parts, strong winds, or weak earthquakes is possible to cause misalignment. Building sway leads to pointing errors at the beam steering that degrades system performance. Hence, outdoor OWC needs a tracking system to maintain accurate beam alignment.

*E. Safety and Regulations*

Exposure to light may cause injury to both the skin and the eye, but the damage to the eye is far more significant because the eye can concentrate the energy of light, covering the wavelengths from around 0.4~1.4um, on the retina. The light with other wavelengths is filtered out by the front part of the eye (i.e., the cornea) before the energy is concentrated. In addition, invisible light at the near-IR band also imposes a safety hazard to human eyes if operated incorrectly. Consequently, the design, deployment, and operation of optical communication systems must ensure that the optical radiation is safe to avoid any injury to the people who might contact it.

## VI. CONCLUSIONS

This paper presented a novel concept Full-Spectrum Wireless Communications for the upcoming 6G cellular systems. Taking advantage of all communication-friendly frequency bands, including radio, microwave, millimeter wave, terahertz, and lightwave (i.e., infrared, visible light, and ultraviolet), FSWC can not only provide sufficient bandwidth to realize extremely high performance of 6G but also enable new paradigm such as integrated communications and sensing.